\documentclass[11pt,fleqn]{article} 

\usepackage[utf8]{inputenc} 

\usepackage{fullpage} 
\usepackage{graphicx,color}
\usepackage{bm}
\usepackage{bbm}
\usepackage{amsmath,amssymb,amsfonts,amsgen,amsbsy,units}
\usepackage{booktabs} 
\usepackage{array} 
\usepackage{paralist} 
\usepackage{verbatim} 
\usepackage{subfig} 
\usepackage{natbib}

\usepackage{hyperref}
\hypersetup{colorlinks=true, linkcolor=blue, citecolor=bluelinkcolor=blue, citecolor=blue}

\providecommand\bnabla{\bm{\nabla}}
\providecommand\der{\mathrm{d}}
\renewcommand{\vec}[1]{\ensuremath{\bm{#1}}} 
\providecommand\bnabla{\gvec{\nabla}}
\renewcommand{\cdot}{\;\vcenter{\hbox{\tiny$\bullet$}}\;} 

\newcommand{\dasr}[1]{{\color{black}#1}}

\newcommand{\aderv}[2]{\frac{\partial {#1}}{\partial {#2}}}
\newcommand{\adervo}[2]{\frac{\der {#1}}{\der {#2}}}

\newcommand{\adervso}[2]{\frac{\der^2 {#1}}{\der {#2}^2}}

\newcommand{\equa}[1]{Eq.~(\ref{#1})}

\newcommand{\equas}[1]{Eqs.~(\ref{#1})}
\newcommand{\equass}[2]{Eqs.~(\ref{#1})--(\ref{#2})}
\newcommand{\equasa}[2]{Eqs.~(\ref{#1}){ }and{ }(\ref{#2})}

\newcommand{\eqn}[2]{\begin{gather}
\displaybreak[2]
#1
\label{#2}
\end{gather}
}

\newcommand{\gat}[2]{\begin{subequations}\label{#2}\begin{gather}
#1
\end{gather}\end{subequations}
}

\title{\bf On the stability of parallel flow in a vertical porous layer with annular cross--section}
\author{{\bf A. Barletta}$^1$\footnote{Email address for correspondence: {\tt antonio.barletta@unibo.it}}\ ,\ {\bf M. Celli}$^1$,\ {\bf D.A.S. Rees}$^2$\\[3mm]
$^1${\small Department of Industrial Engineering, Alma Mater Studiorum Universit\`a di Bologna,}\\
{\small Viale Risorgimento 2, 40136 Bologna, Italy}\\[3mm]
$^2${\small Department of Mechanical Engineering, University of Bath,}\\
{\small Claverton Down, Bath BA2 7AY, United Kingdom}\\
}

\date{} 

\begin{document}
\maketitle

\begin{abstract}\noindent
The linear stability of buoyant parallel flow in a vertical porous layer with \dasr{an} annular cross--section is investigated. The vertical cylindrical boundaries are kept at different uniform temperatures and they are assumed to be impermeable. The emergence of linear instability by convection cells is excluded on the basis of a numerical solution of the linearised governing equations. This result extends to the annular geometry the well--known Gill's theorem regarding the impossibility of convective instability in a vertical porous plane slab whose boundaries are impermeable and isothermal with different temperatures. The extension of Gill's theorem to the annular domain is approached numerically by evaluating the growth rate of normal mode perturbations and showing that its sign is negative, which means asymptotic stability of the basic flow. A concurring argument supporting the absence of linear instability arises from the investigation of cases where the impermeability condition at the vertical boundaries is relaxed and a partial permeability is modelled through Robin boundary conditions for the pressure. With partially permeable boundaries, an instability emerges \dasr{which takes the form of} axisymmetric normal modes. 
\dasr{Then, as the boundary permeability is reduced towards zero, the critical Rayleigh number becomes
infinite.}
\\[5mm]
\end{abstract}

\newpage

\section{Introduction}
In a short paper, \citet{gill1969proof} captured the core thermal property of porous insulating slabs employed for the thermal insulation of buildings. Heat transfer in a vertical plane layer of fluid-saturated porous material with impermeable boundaries having different uniform temperatures is always in a conduction regime, no matter how large is the imposed temperature difference. 
This result is far from being obvious as one assumes \dasr{that a} conduction regime 
\dasr{forms when} the temperature difference is \dasr{sufficiently} small, while cellular convection flow is expected to arise for larger temperature differences. Incidentally, this is precisely what happens if we have a vertical fluid layer instead of a saturated porous medium \citep{vest1969stability}. 

The impossibility of a convective regime, with an enhanced heat transfer rate compared to the conduction regime, means that 
a \dasr{vertical porous slab saturated by air provides a} much more efficient insulation \dasr{than does}
a vertical air gap free of porous material. 
The core of Gill's theorem \citep{gill1969proof} is the linear stability analysis of the basic conduction regime in the vertical porous slab \dasr{in which he used an integral analysis to show that the exponential growth rate of small
disturbances always remains negative.}  This work has offered a fertile ground for further developments; 
\dasr{examples include} the detailed analysis of the growth rate of perturbations \citep{Rees1988, lewisetal1995}, 
\dasr{the inclusions of other effects} \citep{kwokchen1987, Rees2011} and the extension to the nonlinear regime \citep{Stra88}. 
An important feature of Gill's theorem for the absence of thermoconvective instability in a vertical porous slab is that it relies on the hypothesis that the bounding planes are impermeable. If the boundaries are modelled as permeable, then an instability occurs \citep{barletta2015proof}. 

The aim of this paper is to investigate the validity of Gill's theorem when its formulation is adapted to a vertical annular layer of saturated porous material. With reference to thermal insulation techniques, this result may be interesting when we focus, say, on heat transfer from a hot fluid flowing in a vertical round pipe. In fact, insulation of the hot pipe can be done by cladding a low conductivity porous layer around the pipe wall. 

The main difference with respect to the plane slab examined by \citet{gill1969proof} is that the case of an annular porous layer does not allow for a simple rigorous proof of stability. In \dasr{the present} paper we {\dasr have adopted} a strategy based on the numerical evaluation of the perturbation growth rate in order to test the stability of the flow, \dasr{and it is concluded
that the conduction regime is always stable}. 
\dasr{This conclusion}
is further validated by \dasr{considering cases} when the impermeability of the boundaries is made imperfect. 
This imperfection is monitored \dasr{by means of} a dimensionless parameter $\tau$ \dasr{where $\tau=0$ corresponds to
impermeable boundaries}. 
\dasr{It was found that instability in the form of an axisymmetric mode arises whenever $\tau>0$, but that
the critical Rayleigh number becomes infinite as $\tau\rightarrow0$.}

\begin{figure}[t]
\centering
\includegraphics[width=0.3\textwidth]{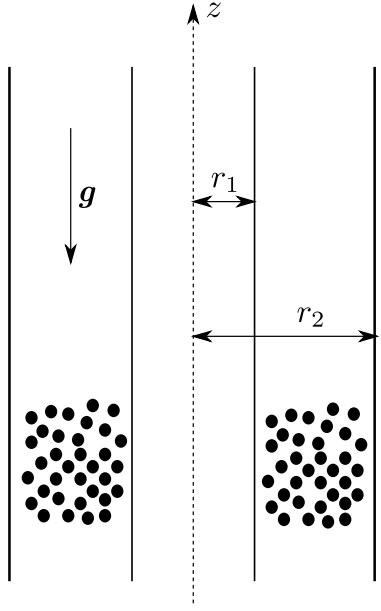}
\caption{Longitudinal cross--section of the cylindrical porous layer}
\label{fig1}
\end{figure}

\section{Mathematical model}
We aim to model a vertical porous layer with cylindrical shape and annular cross-section, bounded by an internal radius $r_1$ and an external radius $r_2$. The radial boundaries $r=r_1$ and $r=r_2$ are impermeable and isothermal with temperatures $T_1$ and $T_2$, respectively. 

\subsection{Governing equations}
By adopting the Oberbeck--Boussinesq approximation and Darcy's law \citep{nield2017convection}, the local balance equations for mass, momentum and energy can be expressed in a dimensionless form as
\gat{
\bnabla \cdot \vec{u} = 0 , \label{1a}\\
\vec{u} = -\, \bnabla p + R\, T \, \vec{e}_z, \label{1b}\\
\aderv{T}{t} + \vec{u} \cdot \bnabla T = \nabla^2 T , \label{1c}
}{1}
with the boundary conditions
\eqn{
r=1 : \qquad u = 0, \quad T = \zeta - 1, \nonumber\\
r=s : \qquad u = 0, \quad T = \zeta. 
}{2}
Here, $\vec{u}=(u,v,w)$ is the seepage velocity with $u$, $v$ and $w$ denoting the radial, angular and axial velocity components, $p$ is the local difference between the pressure and the hydrostatic pressure, $T$ is the temperature, $t$ is time and $\vec{e}_z$ is the unit vector along the axial $z$ axis. The cylindrical coordinates $(r,\phi,z)$ are employed. The dimensionless parameters $R$, $s$ and $\zeta$ are defined as
\eqn{
R = \frac{g \beta (T_2 - T_1) K r_1}{\nu \alpha} , \quad s = \frac{r_2}{r_1}, \quad \zeta = \frac{T_2 - T_0}{T_2 - T_1} ,
}{3}
where $g$ is the modulus of the gravitational acceleration $\vec{g} = - g\, \vec{e}_z$, $\beta$ is the coefficient of thermal expansion, $K$ is the permeability, $\nu$ is the kinematic viscosity, $\alpha$ is the average thermal diffusivity and $T_0$ is the average temperature  employed as the reference value within the Oberbeck--Boussinesq approximation and the definition of the buoyancy force. The scaling employed to exploit the dimensionless equations (\ref{1}) and the boundary conditions (\ref{2}) are given by
\eqn{
\frac{1}{r_1}\ (r, z) \to (r, z),\qquad \frac{\alpha}{\sigma r_1^2}\ t \to t, \qquad \frac{K}{\mu \alpha}\ p \to p,\nonumber\\
\frac{r_1}{\alpha}\ \vec{u} = \frac{r_1}{\alpha}\ (u, v, w) \to (u, v, w) = \vec{u},\qquad  \frac{T - T_0}{T_2 - T_1} \to T,
}{4}
where $\mu$ is the dynamic viscosity and we denoted with $\sigma$ the ratio between the volumetric heat capacity of the saturated porous medium and that of the fluid.

\subsection{Buoyant parallel flow}\label{baflo}
There exists a stationary basic solution of \equasa{1}{2} characterised by a parallel velocity field, a vanishing flow rate and a purely radial temperature gradient. This is given by
\eqn{
u_b = 0, \qquad v_b =0, \qquad w_b = R\, T_b, \qquad p_b = 0, \nonumber\\
T_b = \frac{\ln(r/s)}{\ln(s)} + \zeta, \qquad \zeta = \frac{1}{2 \ln (s)} - \frac{1}{s^2 - 1},
}{5}
where the subscript ``$b$'' means basic state.

\section{Stability of the basic state}
In order to examine whether the basic flow state defined by \equa{5} is stable or not, we first rewrite \equasa{1}{2} according to a pressure--temperature formulation,
\gat{
\nabla^2 p - R\; \aderv{T}{z} = 0, \label{6a}\\
\aderv{T}{t} - \left( \bnabla p - R\, T \, \vec{e}_z \right)\cdot \bnabla T = \nabla^2 T , \label{6b}\\
r=1 : \qquad \aderv{p}{r} = 0, \quad T = \zeta - 1, \nonumber\\
r=s : \qquad \aderv{p}{r} = 0, \quad T = \zeta. \label{6c}
}{6}
Then, we perturb the basic state (\ref{5}) by small amplitude normal modes,
\eqn{
p = p_b + \frac{\varepsilon}{2} \left[ f(r)\, \cos(m\,\phi)\, e^{i(k z - \omega t)}\, e^{\eta t} + \text{c.c.}   \right], \nonumber\\
T = T_b + \frac{\varepsilon}{2} \left[ h(r)\, \cos(m\,\phi)\, e^{i(k z - \omega t)}\, e^{\eta t} + \text{c.c.}   \right],
}{7}
where ``c.c.'' is a shorthand for complex conjugate, $\varepsilon$ is a perturbation parameter such that $|\varepsilon|\ll 1$, $m$ is a non--negative integer, $k$ is the wavenumber, $\omega$ is the angular frequency, $\eta$ is the time growth rate, while $f$ and $h$ are radial amplitude functions. By substituting \equa{7} into \equas{6}, by taking into account \equa{5}, and by neglecting all terms $O(\varepsilon^2)$, we obtain
\gat{
\frac{1}{r}\;\adervo{}{r} \left( r\; \adervo{f}{r} \right) - \left( \frac{m^2}{r^2} + k^2 \right) f - i\,k\, R\, h = 0 , \label{8a}\\
\frac{1}{r}\;\adervo{}{r} \left( r\; \adervo{h}{r} \right) - \left( \frac{m^2}{r^2} + k^2 + i\, k\, R\, T_b + \eta - i\,\omega \right) h + \adervo{f}{r}\; \adervo{T_b}{r} = 0, \label{8b}\\
r=1,s : \qquad \adervo{f}{r} = 0, \quad h = 0. \label{8c}
}{8}
Equations~(\ref{8}) define an eigenvalue problem, where the eigenvalue is the complex quantity $\eta - i\,\omega$, while the pair $(f,h)$ is the eigenfunction. The solution is sought for fixed input values of $(m,k,R,s)$. The real part of the eigenvalue, {\em i.e.} the growth rate $\eta$, is an important parameter as it allows one to detect the linear stability $(\eta \leqslant 0)$ or instability $(\eta > 0)$ of the basic solution. We could not find a rigorous proof that the solution of \equas{8} yields $\eta \le 0$, as for the stability theorem proved by \citet{gill1969proof} for a plane slab. However, there are some results that can be proved and which are the starting point for a numerical solution of \equas{8}.

\subsection{Asymptotic case $R\to 0$}\label{asyR0}
The limit of a vanishing Rayleigh number is a case where no buoyant flow exists and the fluid is isothermal in the basic state (\ref{5}). Then, we expect $\eta \le 0$ as the basic state will be stable. However, it is interesting to determine the value of $\eta$ independently of the information that its sign cannot be positive. By assuming $R=0$, we multiply \equa{8a} by $r\, \bar{f}$, where the bar over the symbol denotes complex conjugation. Then, we integrate by parts over the interval $1 \leqslant r \leqslant s$ by taking into account the boundary conditions (\ref{8c}). The result is
\eqn{
\int_1^s r \left| \adervo{f}{r} \right|^2 \der r + \int_1^s \left( \frac{m^2}{r^2} + k^2 \right) r\, |f|^2 \, \der r = 0 .
}{9}
Equation~(\ref{9}) can be satisfied only with $\der f/ \der r = 0$, if $m=0$ and $k=0$, or with $f=0$ in every other conditions. In either cases, the last term on the left hand side of \equa{8b} is zero. Thus, if we multiply \equa{8b} by $r\, \bar{h}$ and we integrate by parts over the interval $1 \leqslant r \leqslant s$ by taking into account the boundary conditions (\ref{8c}), we obtain
\eqn{
\int_1^s r \left| \adervo{h}{r} \right|^2 \der r + \int_1^s \left( \frac{m^2}{r^2} + k^2 \right) r\, |h|^2 \, \der r + (\eta - i\, \omega ) \int_1^s r\, |h|^2 \, \der r= 0 .
}{10}
Given that $h$ cannot be identically zero, \equa{10} implies that $\omega=0$. In addition, we can conclude that \equa{10} can be satisfied only if $\eta < 0$. 

\begin{figure}[t]
\centering
\includegraphics[width=0.5\textwidth]{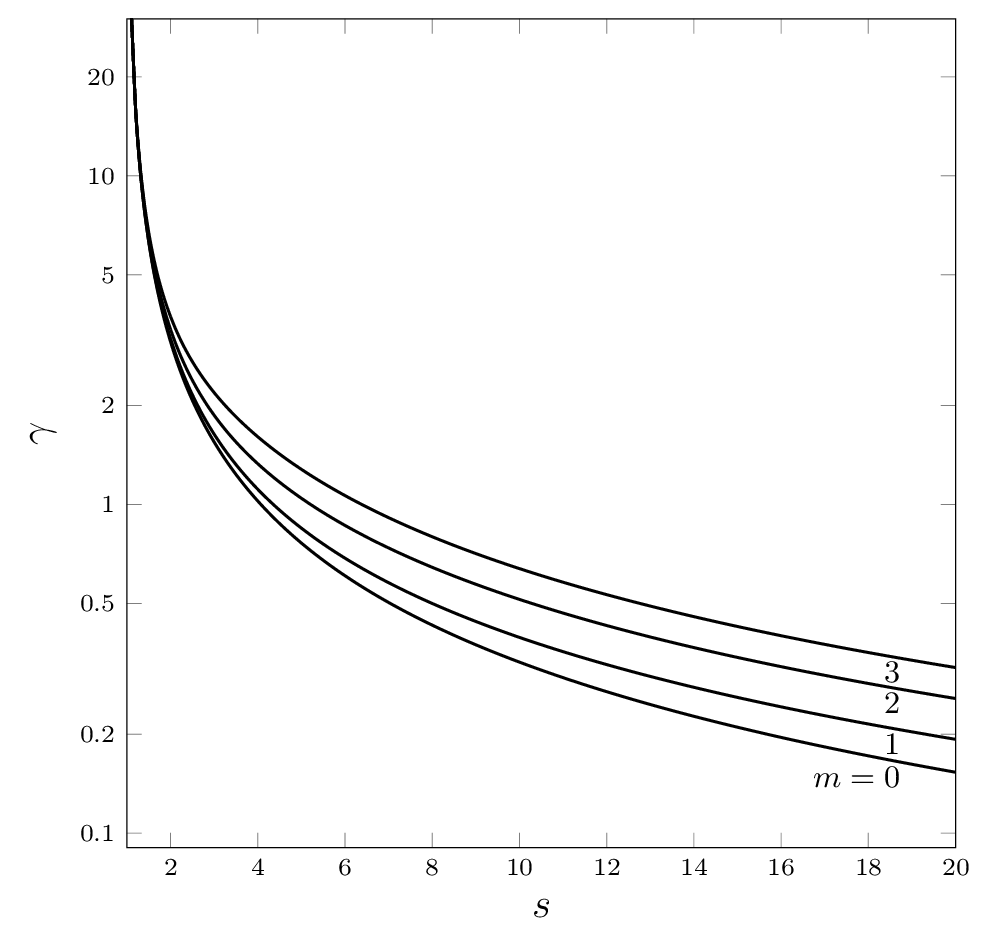}
\caption{Smallest value of $\gamma$ versus $s$ for $m=0,\, 1,\, 2,\, 3$, obtained by \equa{15}}
\label{fig2}
\end{figure}

In order to determine $\eta$ for given $s$ and $k$, we rewrite \equasa{8b}{8c} as
\eqn{
\frac{1}{r}\;\adervo{}{r} \left( r\; \adervo{h}{r} \right) - \left( \frac{m^2}{r^2} + k^2 + \eta \right) h = 0, \nonumber\\
r=1,s : \qquad h = 0.
}{11}
The solution of \equa{11} can be expressed in terms of modified Bessel functions of order $m$, namely
\eqn{
h(r) = C \left[ {\rm K}_m\!\left(\sqrt{k^2+\eta }\right) {\rm I}_m\!\left(r \sqrt{k^2+\eta }\right) - {\rm I}_m\!\left(\sqrt{k^2+\eta }\right) {\rm K}_m\!\left(r \sqrt{k^2+\eta }\right) \right] ,
}{12}
provided that the dispersion relation,
\eqn{
{\rm K}_m\!\left(\sqrt{k^2+\eta }\right) {\rm I}_m\!\left(s \sqrt{k^2+\eta }\right) - {\rm I}_m\!\left(\sqrt{k^2+\eta }\right) {\rm K}_m\!\left(s \sqrt{k^2+\eta }\right) = 0,
}{13}
is satisfied. Here, $C$ is an arbitrary constant, while ${\rm I}_m$ and ${\rm K}_m$ are the modified Bessel functions of first and second kind, respectively. Thus, by employing the properties of Bessel functions (see, for instance, chapter 10 of \citealp{thompson2011nist}), we can evaluate the growth rate $\eta$ as
\eqn{
\eta = -\, k^2 - \gamma^2,
}{14}
where $\gamma$ is a root of
\eqn{
{\rm Y}_m(\gamma)\, {\rm J}_m(s \gamma) - {\rm J}_m(\gamma)\, {\rm Y}_m(s \gamma) = 0.
}{15}
For each choice of $(m,s)$, we are interested in detecting through \equasa{14}{15} the smallest $\gamma$ as it yields the less stable condition, {\em i.e.} that where $\eta$ is at its largest.
Plots of the value of $\gamma$ versus $s$ for $m=0$ to $3$ are reported in Fig.~\ref{fig2}. We can draw some afterthoughts: the growth rate $\eta$ for $R=0$ is always strictly negative; the value of $\eta$ decreases with $m$ and $k$ while it increases with $s$. The former feature means that the basic state is asymptotically stable, according to our linear analysis. However, the most important fact is 
\dasr{that \equa{14} provides}
an accurate starting point for the computation of the eigenvalue $\eta - i\,\omega$ when the Rayleigh number gradually increases 
\dasr{from} zero, for fixed $(m,k,s)$.

\subsection{Asymptotic case $k\to 0$}\label{asyk0}
The case of an infinitely large wavelength, or an infinitely small wavenumber, can be tackled by a treatment similar to that presented in Section~\ref{asyR0}. In fact, since $R$ appears in \equas{8} only through the expression $k\,R$, the limit $k \to 0$ is just a sub--case of that analysed in Section~\ref{asyR0}. Then, we can just use the results drawn in that section and specialise them by setting a zero wavenumber. In physical terms, we infer that axially invariant, or $z$ independent, modes cannot activate the instability for every value of $R$.

\subsection{Computation of the eigenvalue}
The solution of the differential eigenvalue problem (\ref{8}) can be approached numerically by employing the shooting method (see, for instance, chapter 9 of \citealp{straughan2008stability}, or chapter 10 of \citealp{barletta2019routes}). The main stages of the numerical procedure are the following:
\begin{enumerate}
\item We solve \equasa{8a}{8b} as an initial value problem where the initial conditions are those imposed at $r=1$. They are given by \equa{8c}. However, as they stand, they are insufficient to match the differential order of \equasa{8a}{8b}. Thus, we need two additional initial conditions. One of them relies on the scale invariance of the homogeneous problem (\ref{8}), which can be broken by imposing $\der h/\der r = 1$ at $r=1$. The second one is the statement $f(1)=\xi_1 + i\,\xi_2$ which does not imply any loss of generality inasmuch as $\xi_1$ and $\xi_2$ are general real parameters.
\item We determine the eigenvalue $\eta - i\, \omega$ together with the parameters $\xi_1$ and $\xi_2$ by employing a root finding algorithm, such as the Newton--Raphson method, applied to the target conditions expressed by \equa{8c}, at $r=s$, namely $\der f/\der r = 0$ and $h = 0$. In fact, since the eigenfunctions $f$ and $h$ are complex--valued, such target conditions entail four different real constraints.
\end{enumerate}
Both stages are implemented by employing the {\sl Mathematica} software (\copyright{} Wolfram Research) and, in particular, we use the built--in functions {\tt NDSolve} and {\tt FindRoot}.

The shooting method can be utilised for every assignment of the input data $(m,k,R,s)$. The analysis carried out in Section~\ref{asyR0} establishes the values of $\eta$, $\omega$, $\xi_1$ and $\xi_2$ for $R=0$,
\eqn{
\eta = -\, k^2 - \gamma^2,\quad \omega = 0,\quad \xi_1 =0, \quad \xi_2 = 0.
}{16}
For every assigned $(m, k, s)$, we gradually increase the value of $R$ above zero step--by--step. At each step, we call the root finding algorithm by initialising the search routine with the solution data found at the previous step. The step size for the increment of $R$ is dynamically adapted according to the extent of the eigenvalue change.

\begin{figure}[t!!]
\centering
\includegraphics[width=\textwidth]{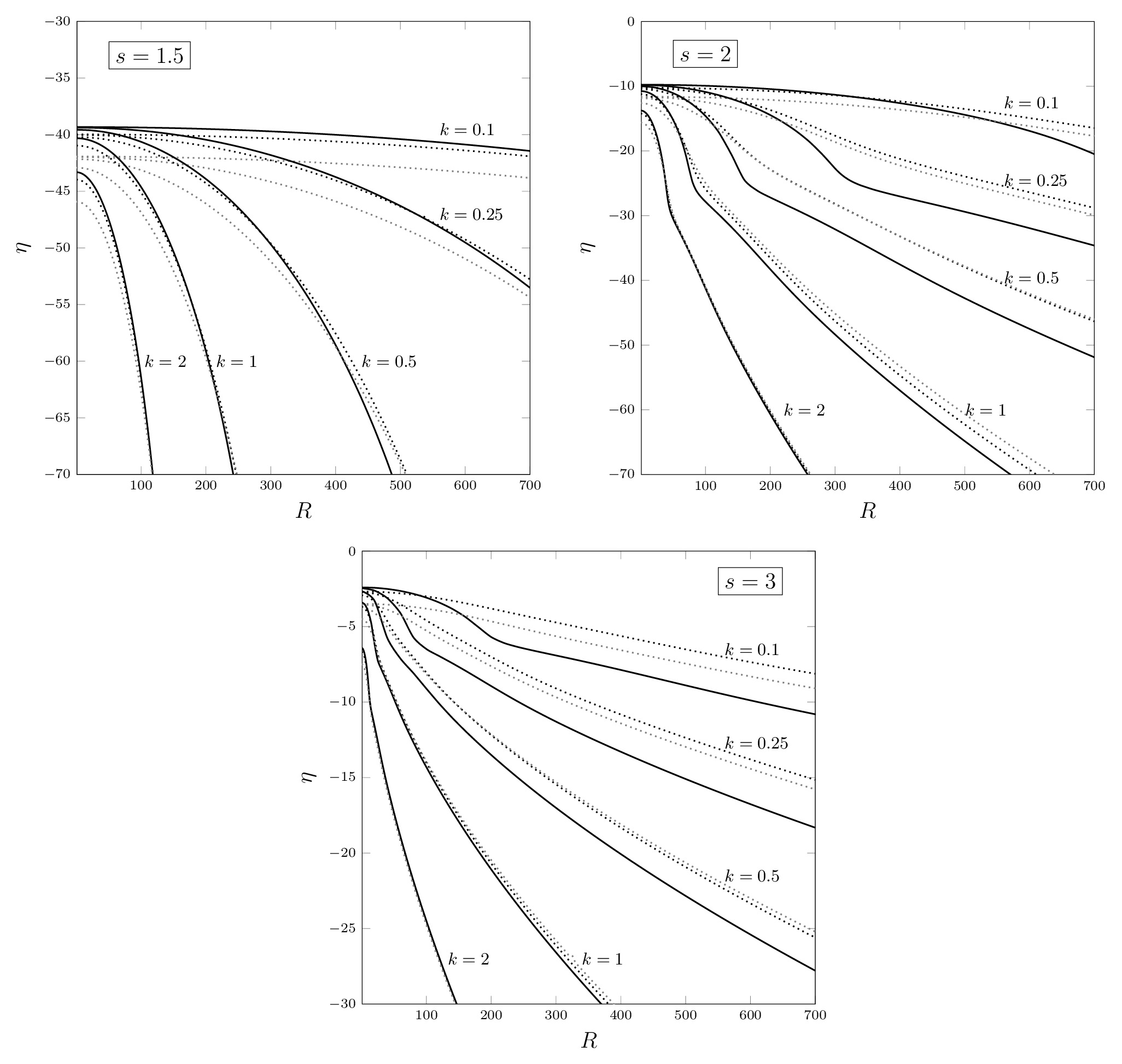}
\caption{Growth rate $\eta$ versus $R$ for different $k$ and either $s=1.5$, $2$ or $3$. Solid lines are for $m=0$, black dotted lines are for $m=1$, while gray dotted lines are for $m=2$}
\label{fig3}
\end{figure}

\subsection{Analysis of the growth rate}\label{grora}
The numerical computation of the complex eigenvalue $\eta - i\, \omega$, for assigned $s$, $k$, $m$ and $R$, allows one to track the growth rate of the normal modes. The sign of the parameter $\eta$ yields the most important information regarding the stable/unstable behaviour of the system. The numerical data collected by varying the governing parameters support the conclusion that $\eta < 0$ in every case. Thus, we reach the conclusion that no instability is possible, exactly as in the case of the plane slab examined by \citet{gill1969proof}.

Figure~\ref{fig3} displays the growth rate $\eta$ versus $R$ for three different aspect ratios, $s=1.5$, $2$ and $3$. The numerical data are relative to $m=0$, $1$ and $2$. These three types of normal mode are identified by solid lines, black dotted lines and grey dotted lines, respectively. There are some distinctive features that may be pinpointed at a glance. The growth rate $\eta$ is a negative, monotonically decreasing, function of $R$ for all the values of $s$, $k$ and $m$ considered in Fig.~\ref{fig3}. The axisymmetric modes  $(m=0)$ yield the less stable conditions, {\em i.e.} those leading to the largest growth rate, only when $R$ is sufficiently small. At larger \dasr{values of} $R$, the $m=1$ or the $m=2$ modes prevail. The selection of the less stable modes largely depends on the value of $k$.  We note that, with $k=0.1$, $\eta$ displays a weak dependence on $k$. This is expected as in Section~\ref{asyk0} we pointed out that $\eta$ is independent of $R$ when $k\to 0$. On the other hand, $\eta$ is poorly influenced by the value of $m$ when $k$ is larger as it becomes apparent in Fig.~\ref{fig3} with $k=2$.

All the data reported in Fig.~\ref{fig3} show that $\eta < 0$ and that the growth rate is a decreasing function of $R$ and of $k$, while the influence of the angular number changes when $k$ increases. As anticipated, the physical information gathered from this analysis is that no linear perturbation mode grows in time, so that the basic flow is always linearly stable.

\section{Changing the impermeability condition at the boundaries}
\dasr{One method of validating the conclusion that the basic flow (\ref{5}) is stable is by relaxing the impermeability
boundary conditions at the inner and outer radii of the cylinder and then investigating the limit of zero
permeability.}
This \dasr{may} be \dasr{achieved} by replacing \equa{8c} by
\eqn{
r=1 : \qquad \adervo{f}{r} - \tau \, f = 0, \quad h = 0, \nonumber\\
r=s : \qquad \adervo{f}{r} + \tau \, f = 0, \quad h = 0. 
}{17}
The parameter $\tau$, \dasr{which is} assumed \dasr{to be} non--negative, marks the departure from impermeability. Obviously, the boundaries \dasr{return to being}
perfectly impermeable only when $\tau \to 0$, a limit where \equasa{8c}{17} coincide. 
Equation~(\ref{17}) \dasr{follows} with Robin boundary conditions for the pressure. Such conditions physically mean that the normal component of the seepage velocity at the boundary is proportional to the pressure difference between the boundary and the external environment.

\dasr{When $\tau\ne0$ our computations show that}
the growth rate $\eta$ may undergo a transition from negative to positive when $R$ becomes sufficiently large, with a neutral stability curve $(\eta = 0)$ 
\dasr{delimiting} the boundary between the \dasr{regions of} parametric stability and instability. 

\begin{figure}[t!!]
\centering
\includegraphics[width=\textwidth]{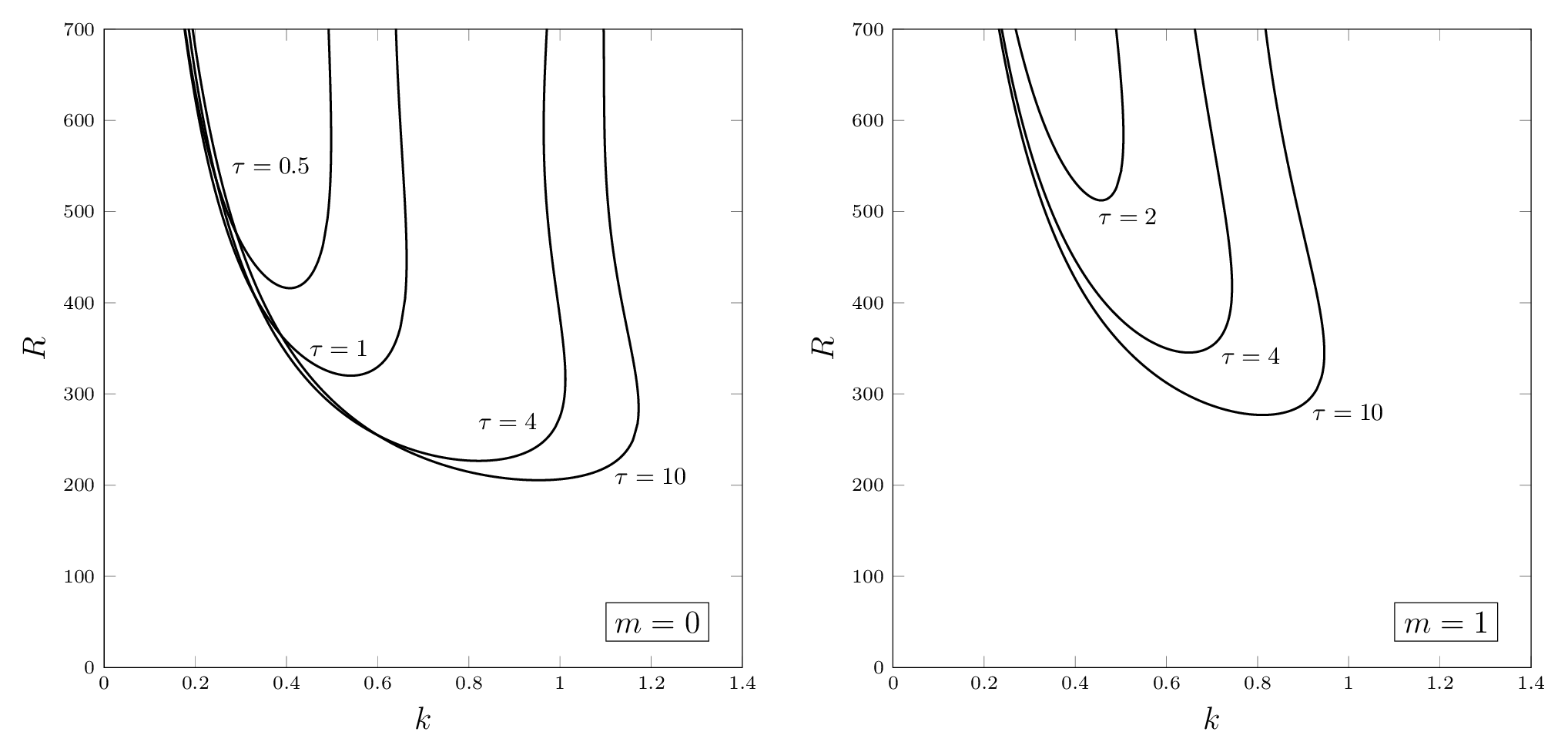}
\caption{Neutral stability curves for the boundary conditions (\ref{17}) with $s=2$, different values of $\tau$ and either $m=0$ or $m=1$}
\label{fig4}
\end{figure}

This behaviour is displayed in Fig.~\ref{fig4}, where neutral stability curves are drawn for the case $s=2$. It is evident that decreasing the value of $\tau$ starting from $\tau=10$ turns into a stabilization of the flow as the neutral stability curves move up and left in the $(k,R)$ plane. The wide gap, for $m=0$, between the minimum of the $\tau=1$ curve and that of the $\tau=0.5$ curve, if compared to that between the curves for $\tau=10$ and $\tau=4$, suggests a very steep variation of the neutral stability curve as $\tau$ becomes smaller and smaller. Just the same conclusion is drawn by comparing the curves with $\tau=4$ and $\tau=2$, relative to the $m=1$ modes. Figure~\ref{fig4} also indicates that, for a given $\tau$, the $m=0$ modes are those activating the instability first or, equivalently, that such modes are the most unstable. 

\begin{figure}[t!!]
\centering
\includegraphics[width=\textwidth]{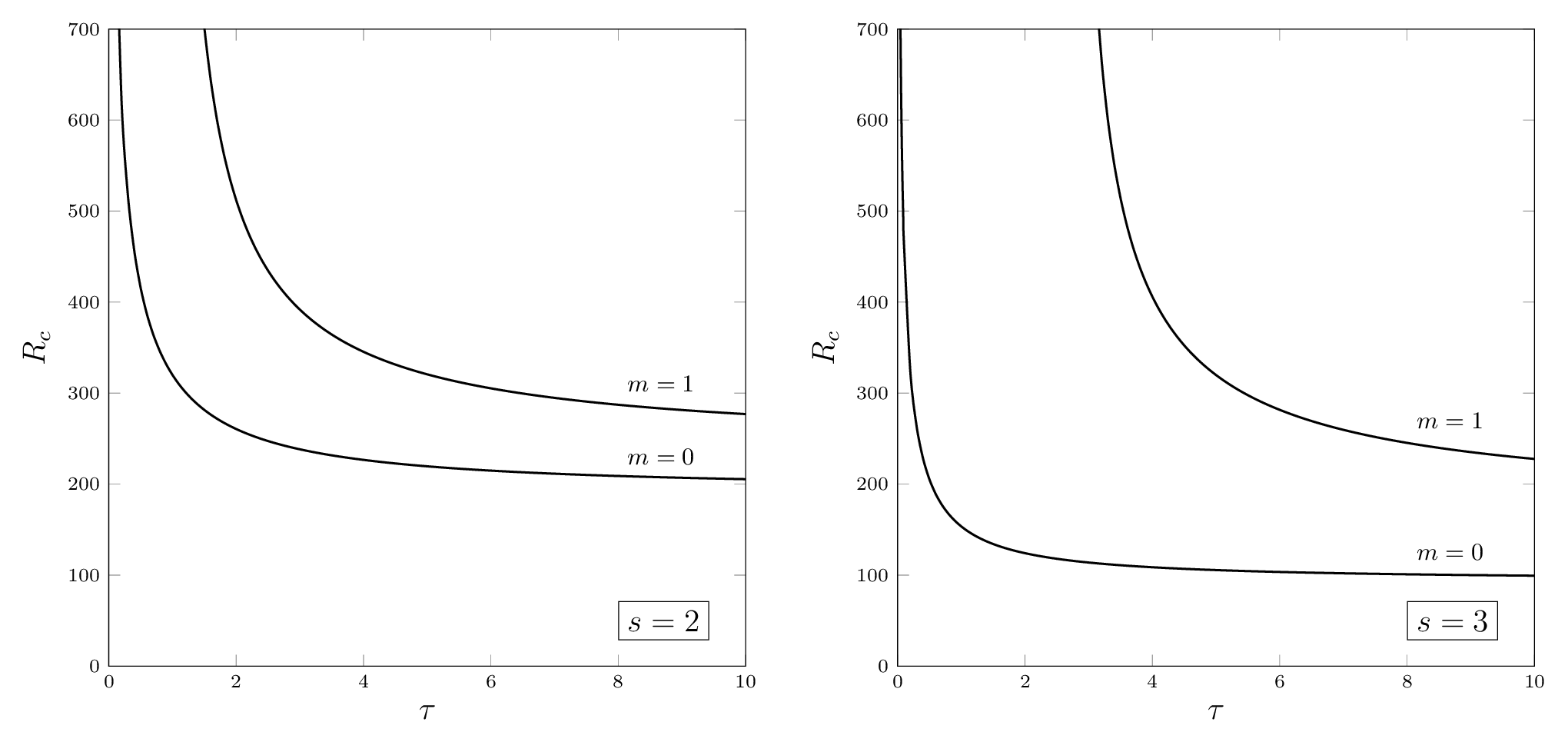}
\caption{Critical value of $R$ versus $\tau$ for the boundary conditions (\ref{17}) and normal modes with either $m=0$ or $m=1$ relative to the cases $s=2$ and $s=3$}
\label{fig5}
\end{figure}

Figure~\ref{fig5} allows one to trace the trend of the critical value $R_c$ versus $\tau$ for two sample cases, corresponding to the ratios $s=2$ and $s=3$. First of all, it must be noted that, in both cases, the most unstable branch is that relative to the axisymmetric normal modes, {\em i.e.} those with $m=0$. The modes with $m=1$ provide a higher branch of instability, while those with $m=2$ yield even larger values of $R_c$ and are not even visible within the vertical range of Fig.~\ref{fig5}. The behaviour reported in Fig.~\ref{fig4}, suggesting a steep 
\dasr{increase in the critical}
value of $R$, is confirmed \dasr{quite clearly} by the curves displayed in Fig.~\ref{fig5}. 
In fact, it is quite evident that the critical value of $R$ tends to infinity when $\tau \to 0$. 
\dasr{In particular, the numerical values for $R_c$ for the case, $s=2$ and $m=0$, satisfy the relation} 
\eqn{
\dasr{R_c \sim 267.671 \tau^{-1/2} + 52.55\tau^{1/2} \qquad\hbox{for}\qquad \tau\ll1.}
}{18}
\dasr{When $\tau=0.05$ this formula yields $R_c=1208.81$ whereas the present numerical calculation gives $R_c=1208.83$.}
\dasr{Such a result shows clearly that $R_c\rightarrow\infty$ as $\tau\rightarrow0$, and it} 
yields an indirect supporting argument for the conclusions drawn in Section~\ref{grora}. The basic buoyant flow described in Section~\ref{baflo} is always linearly stable. We have in fact some further information, as the analysis illustrated in Figs.~\ref{fig4} and \ref{fig5} conveys the awareness that breaking the impermeability condition as it is given by \equa{8c} induces the instability of the basic parallel flow.

\section{Conclusions}
The validity of Gill's theorem \citep{gill1969proof}, originally formulated for a vertical plane porous slab subjected to a boundary temperature difference, is examined for the case of an annular porous layer. More precisely, the stability of the basic stationary conduction regime, implying a vertical buoyant flow through the porous layer, is tested \dasr{by the introduction of} linear normal mode perturbations. Analytical results have been discussed for the special cases where either the Rayleigh number tends to zero, or the wavenumber of the normal modes tends to zero.

In the absence of a general rigorous proof of stability, a numerical solution method has been employed to compute the growth rate of the perturbation modes. In all the tested cases the growth rate turned out to be negative, meaning that the basic flow state is always stable. An additional argument supporting this conclusion has been provided by examining the effects of relaxing the impermeability boundary conditions. Robin pressure conditions, 
\dasr{as mediated}
by the dimensionless parameter $\tau$, are introduced instead of Neumann conditions. This change turned out to activate an instability which, however, disappears in the limiting case where the Robin conditions \dasr{for the pressure} tend towards the Neumann boundary conditions 
\dasr{that correspond to impermeable boundaries.}

\section*{Acknowledgements}
A. Barletta and M. Celli acknowledge financial support from the grant PRIN 2017F7KZWS provided by the Italian Ministry of Education and Scientific Research.

%

\begin{thebibliography}{}

\bibitem[\protect\astroncite{Barletta}{2015}]{barletta2015proof}
Barletta, A., A proof that convection in a porous vertical slab may be
  unstable.
\newblock {Journal of Fluid Mechanics}, {\bf 770},~273--288 (2015).

\bibitem[\protect\astroncite{Barletta}{2019}]{barletta2019routes}
Barletta, A., {Routes to Absolute Instability in Porous Media}.
\newblock Springer, New York (2019).

\bibitem[\protect\astroncite{Gill}{1969}]{gill1969proof}
Gill, A.~E., A proof that convection in a porous vertical slab is stable.
\newblock {Journal of Fluid Mechanics}, {\bf 35},~545--547 (1969).

\bibitem[\protect\astroncite{Nield and Bejan}{2017}]{nield2017convection}
Nield, D.~A., Bejan, A., {Convection in Porous Media}, 5th edition.
\newblock Springer, New York (2017).

\bibitem[\protect\astroncite{Kwok and Chen}{1987}]{kwokchen1987}
Kwok, L.~P., Chen, C.~F., Stability of thermal convection in a vertical porous layer. 
\newblock {A.S.M.E. J.~Heat Transfer} {\bf 109},~889--893 (1987)

\bibitem[\protect\astroncite{Lewis et~al.}{1995}]{lewisetal1995}
Lewis, S., Bassom, A.~P., Rees, D.~A.~S., The stability of vertical thermal boundary-layer flow in a porous medium.
\newblock {European Journal of Mechanics B Fluids}, {\bf 14},~395--407 (1995).

\bibitem[\protect\astroncite{Olver et~al.}{2010}]{thompson2011nist}
Olver, F. W.~J., Lozier, D.~W., Boisvert, R.~F., Clark, C.~W., {NIST Handbook
  of Mathematical Functions}.
\newblock Cambridge University Press (2010).

\bibitem[\protect\astroncite{Rees}{1988}]{Rees1988}
Rees, D. A.~S., {The stability of {Prandtl-Darcy} convection in a vertical porous layer}.
\newblock {International Journal of Heat and Mass Transfer}, {\bf
  31},~1529--1534 (1988).

\bibitem[\protect\astroncite{Rees}{2011}]{Rees2011}
Rees, D. A.~S., {The effect of Local Thermal Nonequilibrium on the stability of convection in a vertical porous channel}.
\newblock {Transport in Porous Media}, {\bf 87},~459--464 (2011). 

\bibitem[\protect\astroncite{Straughan}{1988}]{Stra88}
Straughan, B., {A nonlinear analysis of convection in a porous vertical slab}.
\newblock {Geophysical {\&} Astrophysical Fluid Dynamics}, {\bf 42},~269--275
  (1988).

\bibitem[\protect\astroncite{Straughan}{2008}]{straughan2008stability}
Straughan, B., {Stability and Wave Motion in Porous Media}.
\newblock Springer, New York (2008).

\bibitem[\protect\astroncite{Vest and Arpaci}{1969}]{vest1969stability}
Vest, C.~M., Arpaci, V.~S., Stability of natural convection in a vertical slot.
\newblock {Journal of Fluid Mechanics}, {\bf 36},~1--15 (1969).

\end{thebibliography}

\end{document}